\documentclass[letterpaper,twocolumn,fleqn]{article} 

\usepackage{comment} 
\usepackage{lipsum} 
\usepackage{fullpage} 
\usepackage{graphicx}
\usepackage{mathtools}
\usepackage{amsmath, mathrsfs}
\usepackage{float}
\usepackage{amsfonts}
\usepackage{units}
\usepackage{bm}
\usepackage{color}
\usepackage{subcaption}
\usepackage{ist}
\DeclareUnicodeCharacter{2212}{-}

\title{Advantage of Machine Learning over Maximum Likelihood in Limited-Angle Low-Photon X-Ray Tomography}

\author{Zhen Guo, Department of Electrical Engineering and Computer Science,
Massachusetts Institute of Technology,
Cambridge, Massachusetts, 02139,
USA; Jung Ki Song, Department of Mechanical Engineering,
Massachusetts Institute of Technology,
Cambridge, Massachusetts, 02139,
USA; George Barbastathis, Singapore-MIT Alliance for Research and Technology (SMART) Centre, Singapore 138002; Michael E. Glinsky, Courtenay T. Vaughan, Kurt W. Larson, Sandia National Laboratory
Albuquerque, New Mexico, 87123, USA; Bradley K. Alpert, Applied and Computational Mathematics Division,
National Institute of Standards and Technology,
Boulder, Colorado, 80305, USA; Zachary H. Levine, Quantum Measurement Division,
National Institute of Standards and Technology,
Gaithersburg, Maryland 20899, USA
}

\begin{document} 

\maketitle 

\thispagestyle{empty} 


\begin{abstract}
Limited-angle X-ray tomography reconstruction is an ill-conditioned inverse problem in general. Especially when the projection angles are limited and the measurements are taken in a photon-limited condition, reconstructions from classical algorithms such as filtered backprojection may lose fidelity and acquire artifacts due to the missing-cone problem. To obtain satisfactory reconstruction results, prior assumptions, such as total variation minimization and nonlocal image similarity, are usually incorporated within the reconstruction algorithm. In this work, we introduce deep neural networks to determine and apply a prior distribution in the reconstruction process. Our neural networks learn the prior directly from  synthetic training samples. The neural nets thus obtain a prior distribution that is specific to the class of objects we are interested in reconstructing. In particular, we used deep generative models with 3D convolutional layers and 3D attention layers which are trained on 3D synthetic integrated circuit (IC) data from a model dubbed CircuitFaker. We demonstrate that, when the projection angles and photon budgets are limited, the priors from our deep generative models can dramatically improve the IC reconstruction quality on synthetic data compared with maximum likelihood estimation. Training the deep generative models with synthetic IC data from CircuitFaker illustrates the capabilities of the learned prior from machine learning. We expect that if the process were reproduced with experimental data, the advantage of the machine learning would persist. The advantages of machine learning in limited angle X-ray tomography may further enable applications in low-photon nanoscale imaging.
\end{abstract}

\section{Background}
\label{sec:intro}
Limited-angle X-ray tomography has the ability to image the interior of three-dimensional (3D) objects non-invasively without collecting the measurements from a full set of rotation angles. It has drawn wide attention in practical nanoscale imaging due to its advantage in having a relatively short data acquisition time.  Also, in some cases, not all angles are physically accessible.  

Typically, the tomographic imaging system consists of a sample holder, an objective zone plate, and a CCD camera, and the illumination is generated from the X-ray source.  
The measurements are taken from a series of rotational angles with respect to the sample of interest, where a cone-beam geometry is generally assumed to produce the ray projection from the source point to the sample, and then from sample to the center of each detector pixel. After the measurements, objects subsequently can be reconstructed based on 3D computed tomography algorithms. Theoretically, full-angle measurement is preferred to avoid the missing cone problem in the reconstruction process. In practice, however, limited-angle measurement is often used due to the time of acquiring the full angle measurement. 
For objects or samples that are radiation sensitive, a low-photon budget per scan is also preferred to minimize the total exposure and potential damage.  

Limited-angle X-ray tomography is an ill-conditioned inverse problem~\cite{davison1983ill, rantala2006wavelet}, where the goal is to find a discrete representation of an object $f$ based on the observations $g$ taken on a digital camera at limited number of angles.  When limited-angle tomography is applied where the illumination is limited to only a few photons, the noise sensitivity of poor conditioning becomes even more evident: not only do the collected Fourier slices cover the Fourier space unevenly due to the limited scans, they are also inaccurate due to the presence of shot noise in the measurements. Limited angle X-ray tomography therefore has had limited utility for radiation sensitive samples. To solve the limited angle X-ray tomography with photon scarcity, regularization is required during the reconstruction process.  Some improvement can be obtained by extrapolating missing data~\cite{ferreira1994interpolation}.  Data consistency conditions, e.g.,
the Helgason-Ludwig consistency conditions, further improves the quality of extrapolation~\cite{huang2017restoration}. Still, extrapolating methods are struggling with complex structures and are not robust to noise in the experimental measurements. 

Iterative algorithms with constraints known a priori  reconstruct the object using multiple steps. 
The algorithms start with an assumed object, simulate the measurement from the assumed object, compare the experimental measurements and simulated measurements, and then update the object based on the difference between measurements and simulations. The last step also includes discrepancy in the prior terms into the computation of the update. The process continues until a certain convergence criterion is achieved. Iterative algorithms with prior constraints such as total variation minimization and nonlocal image similarity often exhibit improved resilience to noise. Previously, we also have shown a Bayesian approach~\cite{Levine2019} based on the Bouman-Sauer formulation for the iterative reconstruction algorithm~\cite{Bouman1993}. The regularization constrains the iterative optimization to a subdomain in which the object belongs, thereby effectively improving the reconstruction quality~\cite{sato1981tomographic,verhoeven1993limited}. 

Recently, machine learning has been  applied successfully to limited-angle tomography to overcome the challenge in solving the inverse problem. Efforts have been made in using learned priors to provide information of missing data during reconstruction~\cite{huang2019traditional, wurfl2018deep, huang2020limited}. In particular, deep learning, a subset of machine learning that is based on artificial neural networks with representation learning, achieved promising results~\cite{huang2020limited, bubba2019learning, huang2019data}. 

For tomographic inverse problems, U-net-like neural network architectures~\cite{ronneberger2015u-net} have been widely used to produce reconstruction pixel-by-pixel. For example, U-net has been used to predict invisible singularities in the image object~\cite{bubba2019learning}, to generate missing projections with a data-consistent reconstruction method~\cite{huang2019data}, and to improve the reconstruction quality from the conventional FBP method~\cite{huang2020limited}. Unlike the general priors in iterative algorithms, the learned prior from a deep learning method is conditioned on a large amount of paired training data. Therefore, the learned prior explores the statistical property of the training distribution.

In this work, we extend the application of limited-angle X-ray tomography to experimental conditions of low-photon incidence. To approximate the resulting ill-conditioned inverse problem and to obtain a satisfactory reconstruction quality, we apply machine learning to determine a prior distribution for the reconstruction process. In particular, we use deep generative models with 3D convolution and 3D attention which are trained on 3D synthetic integrated circuit (IC) data from a model dubbed CircuitFaker. We demonstrate that the priors from our deep generative models  drastically improve the IC reconstruction quality on synthetic data from maximum likelihood estimation when the projection angles and photon budgets are limited. Beyond existing research that uses machine learning for limited angle X-ray tomography, our generative model exhibits improvements over maximum likelihood reconstructions under low-photon conditions. 
We further examine different neural network architectures to reveal some of the important network design choices for solving the inverse problem.

\section{Method}

The overall pipeline of our method is organized as follows: First, we generate synthetic integrated circuit (IC) layouts using CircuitFaker, an in-house model. Then, we simulate the limited-angle X-ray tomography radiographs. 
Next, we apply the maximum likelihood method to generate an initial IC reconstruction. Finally, we feed the initial IC reconstruction to the (trained) deep generative model and compare the reconstruction quality of the test set by calculating the bit error rate.

\subsection{CircuitFaker}
\begin{figure}
    \centering
    \includegraphics[width=80mm]{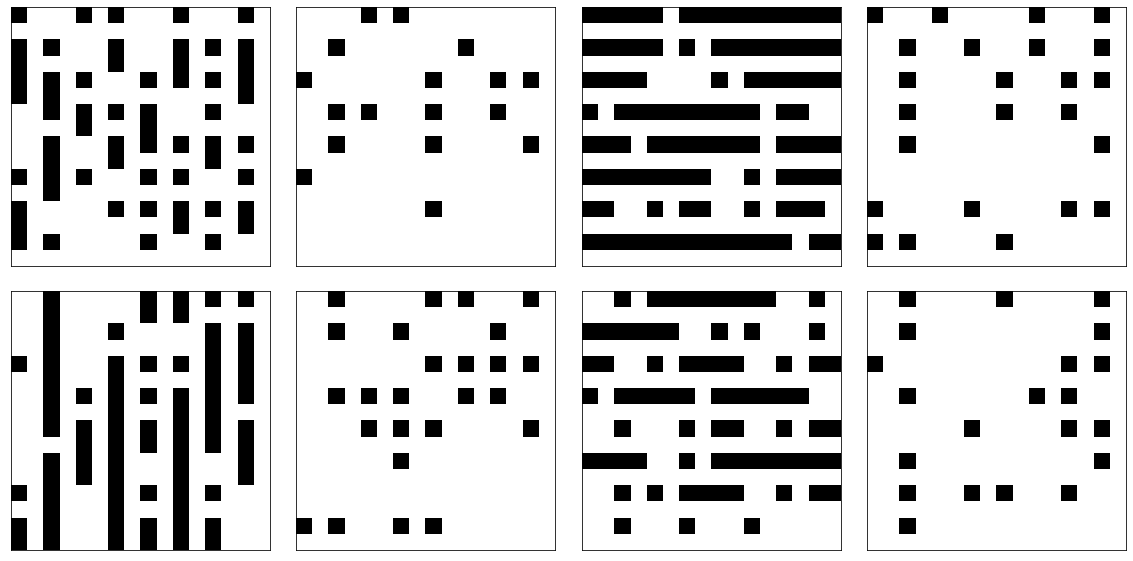}
    \caption{Selected $16\times16\times8$ circuit from CircuitFaker. Each image is a slice of 2D layer in the $z$ dimension. The value of $z$ increases as a raster scan of the 8 slices shown.  Black indicates copper and white indicates silicon. Here, $x$ layers are the first and fifth layers in $z$, $y$ layers are the third and seventh layers in $z$. Others are via layers.}
    \label{fig:expcircuit}
\end{figure}

CircuitFaker is an algorithm that generates a random set of voxels with binary values which resembles an integrated circuit interconnect. The synthetic circuits from CircuitFaker is the class of artificial objects for tomographic reconstruction, and the implicit correlations in their spatial features are the priors to be assumed or to be learned for the inverse algorithms. A particular draw of CircuitFaker assigns a bit in each of $N = N_x N_y N_z$ locations.
These locations are indexed as $i_{\ell} = 1, ..., N_{\ell}$, with $\ell$ = $1$, $2$, $3$ for $x$, $y$, and $z$. All bits are initialized to $0$. In the first round, there are wire seed points for all locations ($i_1$, $i_2$, $i_3$) with $i_1$, ..., $i_3$ odd. Each seed point is set by a Bernoulli draw with probability $p_w$ of getting a $1$.
There are three kinds of layers, $x$, $y$, and via layers. The $x$ wiring layers have index $i_3 = 1$ mod~4. The $y$ wiring layers have $i_3 = 3$ mod~4. The via layers are the others, i.e., $i_3$ even.
In the second round, a point on an $x$ wiring layer to the immediate right of a point with value~1 is set to~1 with probability $p_x$.  A point on a $y$ wiring layer immediately above in plan view a point with a value~1 is set to~1 with probability $p_y$.  Similarly, a point on a via layer immediately above a point with a value 1 is set to~1 with probability $p_z$.
In this paper, we chose these parameters: $N_x=N_y=16$, $N_z=8$, $p_w=0.75$, $p_x=p_y=0.8$, and $p_z=0.5$. Fig.~\ref{fig:expcircuit} shows one of the generated circuits with size $16\times16\times8$. 

\subsection{Imaging geometry for X-ray tomography}

We assumed that each voxel in the circuit is in size of size 0.15~$\mu$m $\times$ 0.15~$\mu$m $\times$ 0.30~$\mu$m. Therefore, the total volume of the circuit is 2.4~$\mu$m $\times$ 2.4~$\mu$m $\times$ 2.4~$\mu$m.  The detector is assumed to be in the $x$-$z$ plane at a tilt angle of $\varphi=0^\circ$.  The rotation axis is $z$.  The detector is 13.44~mm $\times$ 13.44~mm with 32 $\times$ 32 pixels of size 420~$\mu$m $\times$ 420~$\mu$m.  The system operates with a geometric magnification of 5000, with a source-sample distance of 10~$\mu$m.
There are 8 tilt angles from $-30^\circ$ to $+22.5^\circ$ with an increment of $7.5^\circ$.  There is a single source point with a cone-beam geometry.  A single ray is taken from the source point to the center of each detector pixel. Small corrections for variations in the source-to-detector pixel distance, the obliquity, and the source's Heel effect are neglected. The exact detection is written as: 
\begin{equation}
    g^{(0)} = \int \! dE \, D(E) \, I^{(0)}(E) e^{ -\alpha(E)  Af}.
\end{equation}
Here, $A$ is the system matrix, i.e., the distance each projection makes from the source to a detector pixel for all the angles, $f$ is the vector of the object, $E$ is the photon energy, $\alpha(E)$ is the absorption coefficient at energy $E$ for copper, $I^{(0)}(E)$ is the source intensity, $D(E)$ is the detector efficiency. $g^{(0)}$ is the vector of expected measurements on the camera in number of photons for all the angles. With the consideration of Poisson statistics, the final measurements $g$ is:
\begin{equation}
    g \sim \mathscr{P}(g^{(0)})
\end{equation}
where $\mathscr{P}(\lambda)$ denotes Poisson sampling with parameter $\lambda$. In this work, we do not use scatter corrections and we are restricted to a single material, namely copper, at its bulk density of 8.960~g/cm$^3$.  The spectrum consists of two equally weighted lines at 9362~eV and 9442~eV, the Pt L$_\alpha$ fluorescence lines.  The attenuation per voxel is about {2~\%} if copper is present.  The exact value depends on the details of how a ray intersects a voxel.  The specific example was chosen to support an experimental project to perform integrated circuit tomography with a laboratory scale instrument~\cite{Szypryt2021}.

\subsection{Maximum likelihood estimation}
Maximum likelihood estimation as applied to projective tomography is well-known. The method produces the maximum log-likelihood reconstruction circuit $\Tilde{f}$ with a given set of tomographic measurements $g$ (in the number of photon counts for each detector pixel) at different angles. $f$ is the voxelized binary matrix representing the 3D circuit. A voxel value of 1 indicates the presence of copper and 0 indicates silicon. The objective is to find an optimal $\Tilde{f}$ given the measurements g:
\begin{eqnarray}
\label{eq:optimization}
\Tilde{f}(g) &=& \arg\max_{f^{(0)}} \left[ L_{\rm ML}(g | f^{(0)}) + \Psi(f^{(0)})\right] \\
     L_{\rm ML}(g | f^{(0)}) &=&
 \sum_{i} \left[ \ln g_i! - g_i \ln g_i^{(0)} + g_i^{(0)} \right].
\end{eqnarray}
Here $L_{\rm ML}$ is the log-likelihood under the assumption of Poisson statistics, $\Psi$ is a regularization function or log of the Bayesian prior, $g^{(0)}$ is the simulated measurement from a proposed object $f^{(0)}$, $\sum_i$ sums over all individual measurements, and $\Tilde{f}$ is the optimal reconstruction based on maximizing the log likelihood~\cite{Levine2019,Bouman1993}.



\subsection{Bit error rate calculation}
The bit error rate (BER) is introduced as an evaluation metric to assess the performance of the algorithms. It provides a measure of the frequency of misclassification for binary values in the voxels in a given circuit. That is, BER is the probability of classifying a specific voxel in a circuit to be 1 while the ground truth value for the corresponding voxel is actually 0 and vice versa. The procedure for computing bit error rate in this paper is slightly modified from the standard used in communication theory, and is as follows: 
\begin{enumerate}
    \item compute posterior distributions $p(f\!=\!0\,|\,\Tilde{f})$
by multiplying the probability density functions (PDFs) $p(\Tilde{f}\,|\,f\!=\!0)$ and $p(\Tilde{f}\,|\,f\!=\!1)$ and their corresponding prior distributions ($p_0$ and $p_1$). Here, $f\!=\!0$ is the set of silicon voxels in the circuit, $f\!=\!1$ is the set of copper voxels in the circuit;
\item apply a chosen threshold to classify 0 and 1 based on likelihood functions; 
\item compute the error rates for 0 and 1 ($\eta_0$ and $\eta_1$, respectively) by summing over the misclassified region in the probability density functions; 
\item derive an average bit error rate:
$\eta_{\rm avg}=\eta_0 p_0+\eta_1 p_1$.
\end{enumerate}

\subsection{Deep generative models}
A deep generative model, known as a conditional generative adversarial network (cGAN)~\cite{mirza2014conditional}, is a supervised machine learning technique that we used to improve the 3D IC reconstruction over maximum likelihood estimation. When the projection angles and photons per ray are limited, the reconstructions from maximum likelihood estimation contain artifacts due to the missing cone problem and photon shot noise. The bit error rate of these reconstructions would eventually drop below the acceptable threshold of single error per sample when the angular range and/or photon per ray of the tomographic measurements decrease. The deep generative model in our paper improves the Maximum Likelihood reconstructions in these situations through inference of what the likeliest true IC object structure is based on the (noisy) observation and, crucially, {\em what it has learned through its training.} We denote the noisy maximum likelihood estimate $\Tilde{f}$ as the Approximant.

In total, four variants of deep generative models have been investigated. Those variants are: deep generative model (baseline), generative model with axial attention, generative model with a scattering representation, and generative axial model with a scattering representation. 

The first generative model applies the learned prior $p(f)$ to the input distribution $p(\Tilde{f}\,|\,f)$, and estimates $p(f\,|\,\Tilde{f})$ using Bayes' theorem~\cite{holden2021bayesian, nolan2021frequentist}, where $p(f)$ describes a distribution of noiseless IC samples.  
That is, the generative model learns the prior using a combination of generator and discriminator, which are trained to compete with each other until they reach the Nash equilibrium. We note the output of the GAN in equilibrium as $\hat{f}$. The entire generative model is formulated as  follows~\cite{mirza2014conditional}:
\begin{equation}\label{eq:cGAN}
\arg \min_{G} \max_{D} \Big(\mathbb{E}_{f} \big[ \log D(f) \big] + \mathbb{E}_{\Tilde{f}} \big[ \log \big(1 - D(G(\Tilde{f}))\big) \big] \Big)
\end{equation}
The generator in our model is a 3D autoencoder that first learns to convert IC to latent space representation using 3D convolutional encoder, and then decodes the representation back to IC. The discriminator is a 3D convolutional classifier that tries to find whether the output from the generator is realistic or not. Both generator and discriminator in our model have convolutional kernels that are spectrally normalized to stabilize the training process~\cite{miyato2018spectral}. This serves as the baseline generative model for our paper. 

Our second variant of generative model is based on axial attention that harvests the contextual information in the input reconstruction. To build such model, we replace the 3D convolution in the encoder of the generator with full axial attention to extract or detect global features in the input reconstructions. 
This technique reduces the computational complexity of non-local attention to $\mathcal{O}(hwzm)$, where $h$, $w$, $z$, are the height, width, and depth of the input features, respectively, $m$ is the local constraint constant~\cite{wang2020axial}. Therefore, a network with axial attention is more efficient than a self-attention network, enabling long-range and global feature learning to overcome the limitation of locality in convolutional networks.

The third and fourth variants of generative models include the wavelet scattering transform~\cite{anden2014deep, mallat2012group} of the IC reconstruction as additional input to the model. A scattering representation can be produced without training, and the produced representation is capable of including features at multiple scales. When combined with the renormalization technique, the generative model would be further conditioned on the scattering representation in generating realistic IC layout. In particular, the wavelet representation, after being fed to a trainable transformation of a fully connected layer, re-scales and re-shifts the normalized feature values from convolution. 
This way, the information about global features of the IC will be embedded into the reconstruction process of the neural network~\cite{de2017modulating}, which may help the network in learning the mapping from noisy reconstruction to noiseless reconstruction.


\begin{figure}
\centering
\begin{subfigure}[b]{0.5\textwidth}
   \includegraphics[width=79mm]{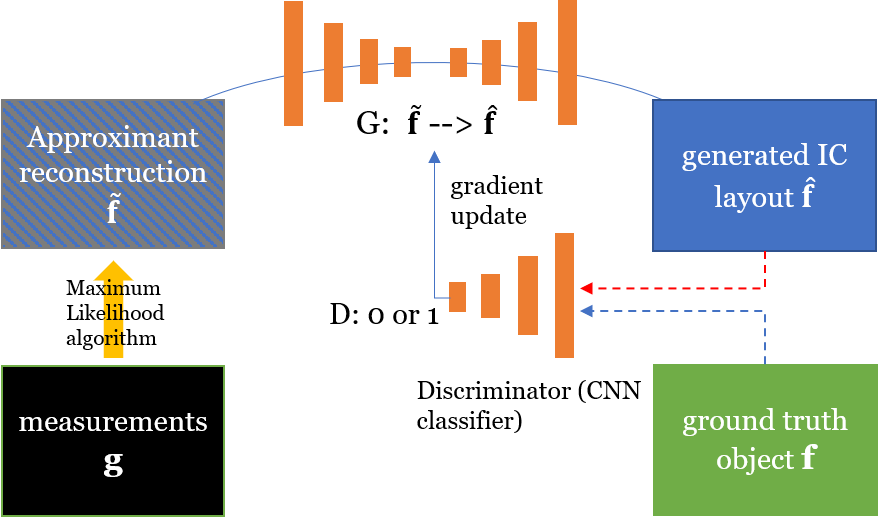}
   \label{fig:Ng1} 
\end{subfigure}

\begin{subfigure}[b]{0.5\textwidth}
   \includegraphics[width=79mm]{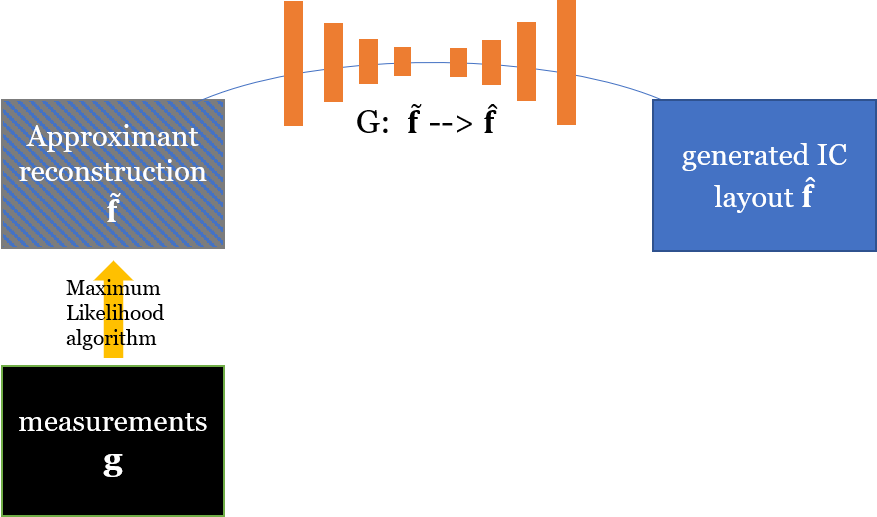}
   \label{fig:Ng2}
\end{subfigure}

\caption[generative framework]{(a) Supervised training of the generative model with pairs of (f, g). (b) Testing/operation on samples never used during training.}
\end{figure}

\subsection{Network training}
Our proposed networks were implemented in Python 3.7.9 using TensorFlow 2.3.1, and trained with an NVIDIA V100 tensor core graphics processing unit on MIT Supercloud~\cite{reuther2018interactive}.  An Adam optimizer~\cite{kingma2014adam} was used with parameters $\beta_1 = 0.9$ and $\beta_2 = 0.999$. Two time-scale update rule (TTUR) technique is used to to stabilize the training of generative network~\cite{radford2015unsupervised, heusel2017gans}, where the initial learning rate was $10^{-4}$ for the generator and $4\times10^{-4}$ for discriminator. In each iteration, generator would be updated four times while discriminator would be updated once.  
Training sets of 1800 reconstructions were generated independently for each condition studied, except that the ground truth was common.
An additional 200 reconstructions per condition were used for testing.
The learning rate would be reduced by half when the validation loss stops improving for 5 epochs. We set the maximum number of iterations to be~200,
and the training would stop early when either the validation loss plateaus for 20 epochs, or the minimal learning rate $10^{-8}$ is reached. This early-stop technique would prevent the model from over-fitting. The loss function for the autoencoder/generator consists of two parts: supervised loss and adversarial loss.  We choose supervised loss to be the negative of the Pearson correlation coefficient $-r_{f, \Tilde{f}}$,
which is defined as 
\begin{equation}
-r_{f, \Tilde{f}} = - \frac{\text{cov}(f, \Tilde{f})}{\sigma_{f} \; \sigma_{\Tilde{f}}},
\end{equation}
where $\text{cov}$ is the covariance, $\sigma$ is the standard deviation.
The total objective of training is to find the optimal generator $G_{\text{opt}}$ given the Approximant $\Tilde{f}$:
\begin{equation}
\begin{split}
    G_{\text{opt}}(\Tilde{f}) & = \mathbb{E}_{f, \Tilde{f}}\big[-r_{f, G(\Tilde{f})} \big]  + \lambda\arg \min_{G} \max_{D} \\
    & \Big(\mathbb{E}_{f} \big[ \log D(f) \big] + \mathbb{E}_{\Tilde{f}} \big[ \log \big(1 - D(G(\Tilde{f}))\big) \big] \Big)
\end{split}
\end{equation}
The hyper-parameter $\lambda$ that controls the degree of generation from input noise to features.

\section{Results}
\begin{figure}
    \centering
    \includegraphics[width=80mm]{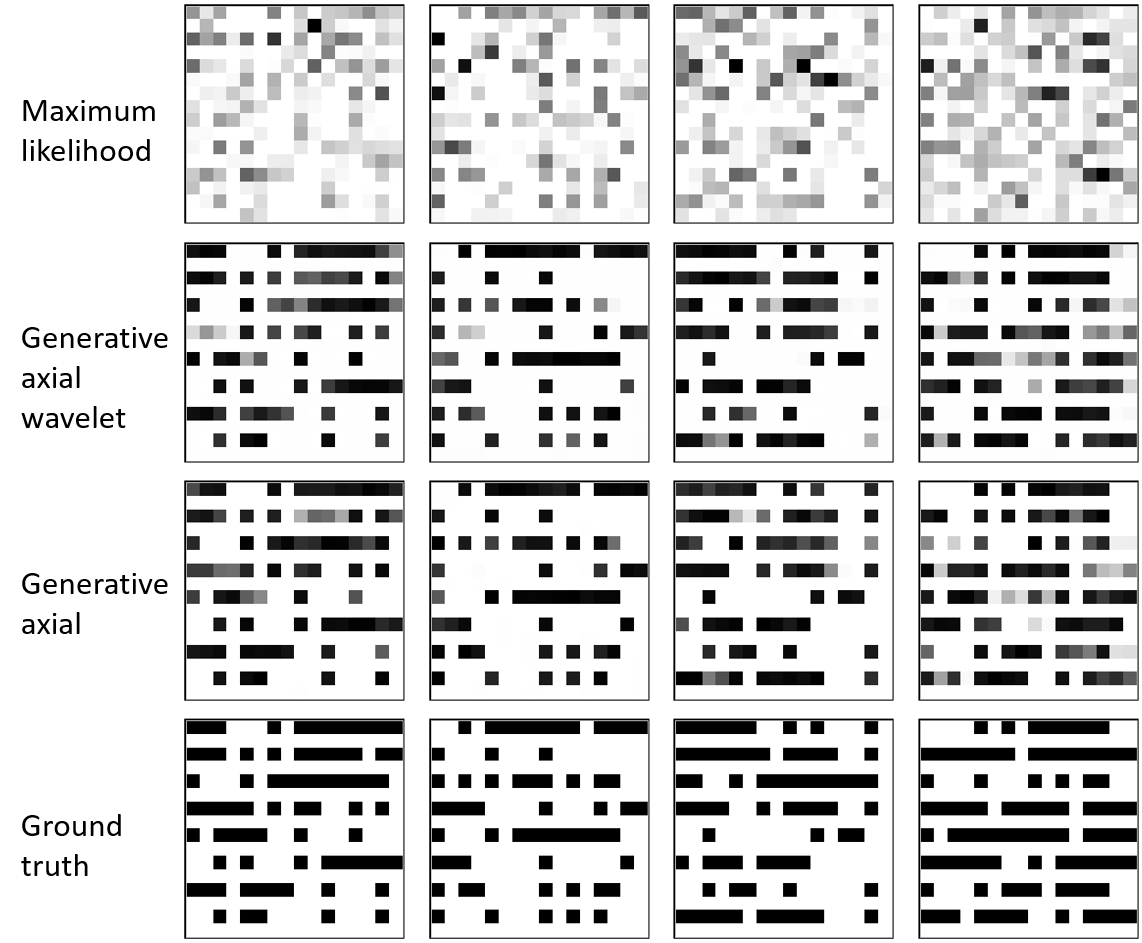}
    \caption{Selected Examples of IC Reconstructions for Different Methods}
    \label{fig:ic-compare}
\end{figure}

Fig.~\ref{fig:ic-compare} shows selected examples of IC reconstructions at limited-angle and low photon conditions. Each row represents different reconstruction methods, and each column represents the same location at the given IC distribution. The last row represents the IC ground truth. Drastic improvement is visible when comparing the maximum likelihood reconstructions and generative reconstructions. 
\begin{figure}
    \centering
    \includegraphics[width=80mm]{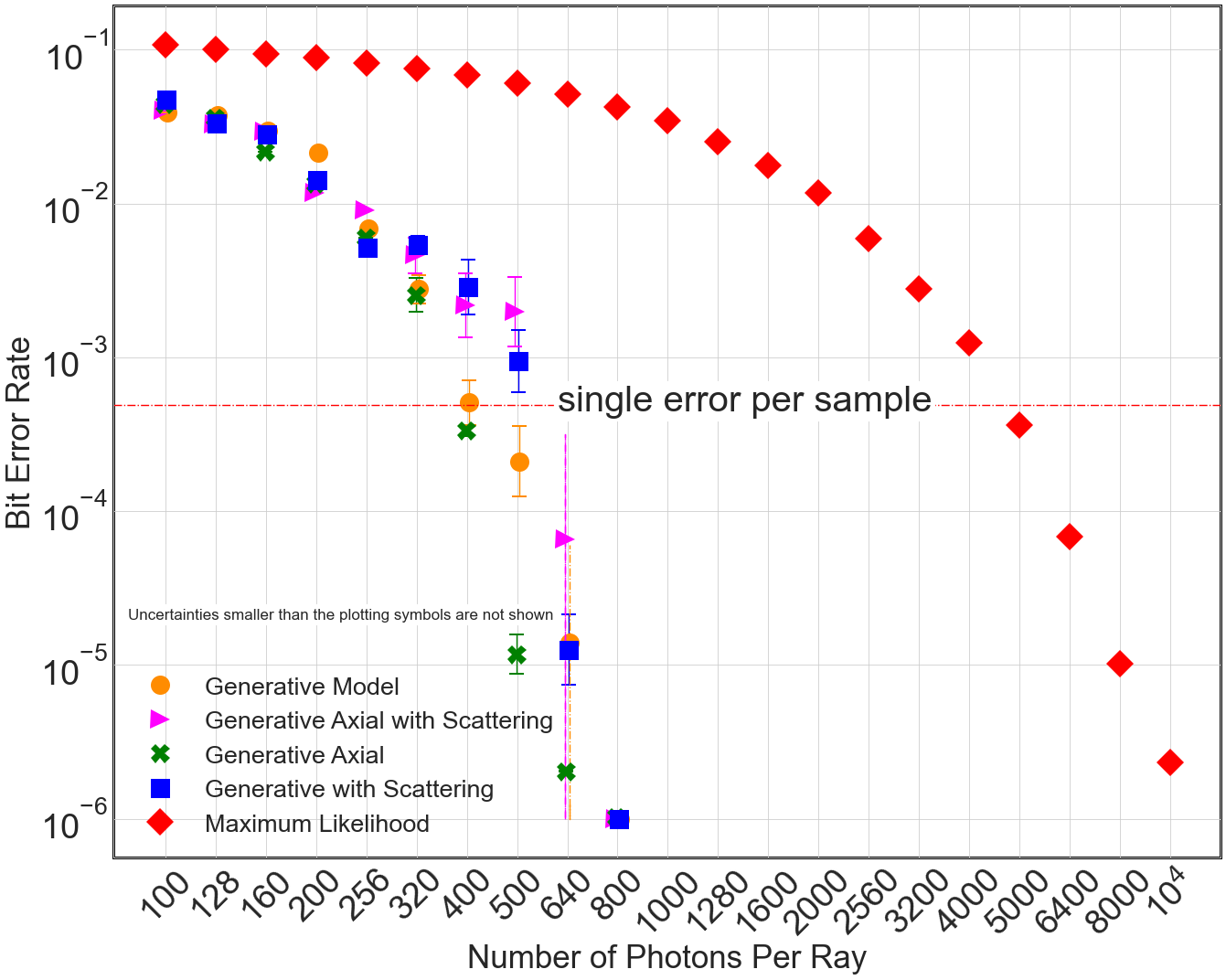}
    \caption{Maximum Likelihood vs. Generative Model Reconstructions with $-30^{\circ}$ to $22.5^\circ$ angular range.}
    \label{fig:MLvsGANb}
\end{figure}
The numerical result of limited angle and low photon tomography is shown in Fig.~\ref{fig:MLvsGANb}. The $x$ axis is the number of photons per ray in the tomographic projection, $y$ axis is the bit error rate of the reconstructed 3D IC samples. The angular range is fixed at $-30^\circ$ to $22.5^{\circ}$ with $7.5^{\circ}$ steps for all. There are five reconstructions we are comparing: reconstruction based on maximum likelihood estimation, and reconstruction based on four variants of generative model. The transition from above a single-error-per-sample to below happens between 320 to~640 photons per ray.  Each of the simulations for the generative model is done with 1800 training sets and 200 test sets, and IC data is $16\times16\times8$ voxels. For the transition cases between 320 and 640 photons per ray, we repeat the simulation with a total of five independent synthetic sets of IC circuits and report the means and standard errors in the plot.  With 640 photons per ray, all the bit error rates from generative model reconstructions drop at least two orders of magnitude relative to the maximum likelihood reconstructions. In particular, the generative model with axial attention performs the best in terms of its lower mean and standard error,  reaching a single error per sample at 400 photons per ray. We may attribute this to the application of axial attention to capture long range interactions within the input. The maximum likelihood reconstructions reach a single error per sample when the number of photons per ray is 5000. Therefore, in simulation, the generative model using machine learning can reduce the photon budget one order of magnitude to reach a single error per sample.

\section{Conclusion}
In this work, we applied machine learning to limited-angle X-ray tomography with low-photon imaging conditions. Our deep-learning model is generative, and it is based on 3D convolutional layers and 3D attention layers, 
and then trained on 3D synthetic integrated circuit (IC) data from CircuitFaker. We demonstrated that the prior distributions from our deep generative models can dramatically improve the IC reconstruction quality on synthetic data compared to a maximum likelihood estimation algorithm when the projection angles and photon budgets are limited. Training the deep generative models with synthetic IC data from CircuitFaker illustrates the capabilities of the prior from machine learning. We expect that if the process was reproduced with experimental data, the advantage of the machine learning would persist. The advantages of using machine learning in limited angle X-ray tomography may further enable applications in low-photon nanoscale imaging.



\section{Acknowledgement}
The authors thank Zachary Grey for extensive comments on an earlier draft.  This research at MIT was supported by IARPA contract number FA8050-17-C-9113. G.B. also acknowledges support by Singapore's National Research Foundation through the Intra-Create grant NRF2019-THE002-0006 ``Retinal Analytics via Machine leaning aiding Physics (RAMP).''  Sandia National Laboratories is a multimission laboratory managed and operated by National Technology and Engineering Solutions of Sandia LLC (NTESS), a wholly owned subsidiary of Honeywell International Inc., for the U.S. Department of Energy's National Nuclear Security Administration (NNSA) under contract DE-NA0003525.  This paper describes objective technical results and analysis. Any subjective views or opinions that might be expressed in the paper do not necessarily represent the views of the U.S. Department of Energy or the United States Government. (Sandia technical report number SAND2021-15957 C).
Research at NIST of B.K.A. and Z.H.L. supported in part by the Office of the 
Director of National Intelligence (ODNI), Intelligence Advanced
Research Projects Activity (IARPA), via agreements D2019-1908080004 and D2019-1906200003. 

Mention of commercial products does not imply endorsement by the authors or their institutions.




\small

\bibliographystyle{ieeetr}
\bibliography{reference}


\begin{biography}
ZG is a graduate student at MIT EECS, who is working on computational imaging and machine learning. GB is a fellow of Optica (formerly known as Optical Society of America) and has been on the MIT faculty since 1999. His interests are in the intersection of computational imaging and machine learning, explicitly embedding imaging physics. ZHL is a Fellow of the American Physical Society and has been a Physicist at NIST since 1994. He led the first team which published a reconstruction of an integrated circuit interconnect.  MEG is physicist at Sandia National Laboratories working on magneto-inertial fusion.  He has worked at LLNL, on advanced ICF concepts and biomedical device design, and in the petroleum and mining industry on the use of machine learning.
\end{biography}

\end{document}